\documentclass[pre,twocolumn,showpacs]{revtex4}
\usepackage{epsfig}

\def\8{\infty}
\def\oh{\frac{1}{2}}

\def\d{\partial}

\def\undertext#1{\vtop{\hbox{#1}\kern 1pt \hrule}}

\def\VEV#1{\left\langle\,#1\,\right\rangle}

\def\dd#1{\frac{d}{d#1}}
\def\dbyd#1#2{\frac{d#1}{d#2}}
\def\pp#1{\frac{\partial}{\partial#1}}
\def\pbyp#1#2{\frac{\partial#1}{\partial#2}}

\def\br{\\ \nonumber & &}

\def\be{\begin{equation}}
\def\ee{\end{equation}}
\def\bea{\begin{eqnarray} & &}
\def\eea{\end{eqnarray}}

\def\rf#1{(\ref{#1})}

\def \PDF{probability distribution function }

\def\rfs#1{Eq.~(\ref{#1})}
\def\Urms {U_{\rm rms}}

\begin{document}

\title{Burgers Equation Revisited}

\author {V. Gurarie}
\affiliation{Department of Physics, Theoretical Physics, Oxford
University, 1 Keble Rd, Oxford OX1 1JP, United Kingdom}

\date{14 March 2002}
%\date{\today}

\begin{abstract}
This paper studies the 1D pressureless turbulence (the Burgers
equation). It shows that reliable numerics in this problem is very
easy to produce if one properly discretizes the Burgers equation.
The numerics it presents confirms the $7/2$ power law proposed for
probability of observing large negative velocity gradients in this
problem. It also suggests that the entire probability function for
the velocity gradients could be universal, perhaps in some
approximate sense. In particular, the probability that the
velocity gradient is negative appears to be $p \approx 0.21 \pm
0.01$ irrespective of the details of the random force. Finally, it
speculates that the theory initially proposed by Polyakov, with a
particular value of the ``anomaly" parameter, may indeed be exact,
at least as far as velocity gradients are concerned.
\end{abstract}


\maketitle

\section{Introduction}

The problem of the randomly driven Burgers equation, or 1D
pressureless turbulence, attracted a lot of attention in the
literature in the last decade
\cite{Yakhot,Polyakov,BMP,GM,Sinai,Frisch,KG,Bec}. Indeed, it is a
tantalizing problem. On the one hand, it incorporates many
features one would expect from a real 3D turbulent fluid. On the
other hand, it is much simpler than the 3D turbulence of the
incompressible fluid, usually studied in the framework of the
Navier-Stokes equation. Its simplicity prompted some researchers
to suggest that it will become the ``Ising model'' of turbulence.
Yet many published papers later, we are still far from a complete
solution to this problem. In fact, many of the aspects of the
Burgers equation remain the subject of an ongoing debate. Several
competing methods exist in the literature, each with its own range
of applicability, which provide different, sometimes mutually
exclusive answers to various aspects of the 1D turbulence.

The situation could be improved if a way existed to check various
statements made about the Burgers equation. Such a check could be
provided by numerical simulations. But the numerical simulations
in turbulence are hard to conduct, they require large expenditure
of resources, and the answers they provide are often too ambiguous
to either confirm or disprove a particular theory. The situation
in Burgers turbulence seems to be no different. Despite several
influential papers on the numerical aspects of Burgers turbulence,
those studies failed to rule out completely or confirm any of the
competing theories currently on the market. The exception to this
seems to be the paper of J. Bec \cite{Bec}, where he confirmed the
asymptotics of a certain probability distribution function
(discussed below) predicted earlier in Ref. \cite{Sinai}.

In this paper, I would like to point out that there exists an
alternative way to study the Burgers equation numerically. The
method involves minimizing the Hopf-Cole functional instead of
trying to solve a complicated nonlinear differential equation.
That turned out to be extremely easy to do numerically. A ten line
program, which can be written in half an hour, produces, after
several minutes of computer time, clean graphs of probability
distribution functions. With the help of this program, I report
confirmation  of some of the existing theories. I also hope that
the method, being so simple, can be used by other researchers to
do a quick check if their proposed theory indeed does not
contradict the numerical experiment.

The rest of this paper is organized as follows. In section II I
formulate the problem of Burgers turbulence and summarize many of
the known facts about it. In section III I show how Burgers
equation must be discretized in order to facilitate doing
numerics. In section IV I discuss the statistics of the velocity
gradients of the Burgers equation using the methods of Lagrangian
trajectories. I also discuss the ``anomaly" hypothesis of Polyakov
and explain its meaning within the Lagrangian approach. In section
V I discuss the numerical simulations, show that they are
compatible with the asymptotic results of Refs.~\cite{Sinai,Bec}
and show that they are also compatible with the Polyakov
hypothesis if the ``anomaly" parameter $\beta$ is chosen to be
equal to $3/2$. Finally in the last section VI I also show that
the compatibility with the known results on the behavior of pinned
charge density waves \cite{AR,Fogler,GC,GC1} also forces the value
$\beta=3/2$. I speculate that since the problem of pinned charge
density waves and the asymptotics of Refs.~\cite{Sinai,Bec} relate
to different properties of Burgers turbulence, there is a chance
that the Polyakov's theory with $\beta=3/2$ is indeed exact, at
least as far as the calculation of probabilities of velocity
gradients is concerned, unless a remarkable and unlikely
coincidence is at play here.

\section{Formulation of the Problem}
The Burgers turbulence problem can be formulated in just a few lines.
Take a pressureless fluid in
1D, introduce its velocity field $u(x,t)$ where $x$ is space and $t$ is time and
write down the Navier-Stokes equation
\be
\label{burgers}
\partial_t u+u~\partial_x u - \nu~\partial^2_x u = f(x,t).
\ee Here $\nu$ is the viscosity of the fluid which is taken to be
very small. $f(x,t)$ is a force acting on the fluid, which is
taken to be random, white noise in time, and a smooth function in
space \be \VEV{f(x,t) f(x,t')}= \delta(t-t') F(x-x'), \ee where
$F(x)$ is a function which is positive at $x=0$ and smoothly goes
to zero when $x \gg L$, where $L$ sets the force length scale. As
far as boundary conditions are concerned, it is usually assumed
that $u$ is either periodic in space (in which case $F$ has to be
periodic as well), or the space is infinite and $u \rightarrow 0$
at infinity. Note that the center of mass motion decouples in both
cases, to give \be \dd{t} \int dx~u =  \int dx~f, \ee which is but
the standard Brownian motion. For this reason one usually selects
the force $f$ to be a derivative of another function $h(x,t)$,
$f=\partial_x h$. With this restriction, the average velocity
vanishes $\VEV{u}=0$ and it makes sense to define the {\sl root
mean square velocity} as $U_{\rm rms} = \sqrt{\VEV{u^2}}$. It
measures the average amplitude of the fluctuation of the velocity
field. A basic assumption in the theory of turbulence is that this
quantity does not depend on the viscosity $\nu$ as it is taken to
zero.  That allows to determine $\Urms$ via simple hydrodynamic
scaling to be $ \Urms \propto  (L F(0))^{\frac{1}{3}}$.

Once the equation \rf{burgers} is written, one would like to calculate various averages and
probabilities. A particular popular question focuses on the probability distribution function
of the velocity at two different points. Define  $P(\sigma)$ as the
probability of  observing a velocity
gradient $\sigma=\partial_x u(x)$ in a particular point in space.
Similarly $P_d(v,r)$ is the probability of observing a difference in velocities
$v=u(x+r)-u(x)$
at two points separated by a distance $r$ from each other.
These quantities received a lot of attention
in the literature, possibly because they are related
to the turbulent advection of particles suspended in the fluid, which is a very important
topic in real 3D turbulence.

The following is a highly subjective historical overview of the progress made in understanding
\rf{burgers}. As always, the choice of highlights is very personal one, and I apologize to the authors
whose contributions are not mentioned here.

The first to address the behavior of $P_d(v)$ were V. Yakhot and A. Chekhlov \cite{Yakhot}, who
argued that it should not be symmetric under $v \rightarrow -v$. They were also the first
to study this function numerically.

Soon thereafter, A. Polyakov \cite{Polyakov} suggested an approach
to calculation of these functions. He suggested an equation that
$P_d(v)$ should satisfy. The drawback of his approach was that it
was not just one equation but a family of equations parametrized
by a parameter (referred to as ``$b$-anomaly'') whose value
remained undetermined. At about the same time, Bouchaud, Mezard
and Parisi \cite{BMP} attacked the problem using replica trick and
the Gaussian variational ansatz. The problem with their approach
was that it was not well suited to computing quantities such as
$P(\sigma)$ and $P_d(v)$.

In subsequent work \cite{GM} of the author with A. Migdal, it was
shown that at least the asymptotics of $P(\sigma)$ and $P_d(v)$ at
$\sigma, v \rightarrow +\infty$ can be derived carefully in the
saddle point approximation. It was shown that \bea \label{RHS}
P(\sigma) \propto \exp \left( - \frac{2 \sigma^3}{3  G} \right), \
\sigma \gg G^{\frac{1}{3}}\approx {\Urms \over L} \br P_d(v)
\propto \exp \left( - \frac{2 v^3}{3 r^3 G} \right), \ v \gg \Urms
\eea where $G=-\partial^2_x F(x)|_{x=0}$. The drawback of the
saddle point approach was that it allowed to determine the right
hand side asymptotics only. But that asymptotics is expected to
depend on whether the force $f$ of \rf{burgers} is Gaussian. The
asymptotics \rf{RHS} is true only if the random force of the
Burgers equation is Gaussian. But it does not have to be. For
example, one could take the random force whose strength is always
less than a certain value, $|f| < f_{\rm max}$. Then the
asymptotics \rf{RHS} would not be expected to hold. But if it is
so, then one could legitimately ask if at least some part of the
functions $P(v)$ and $P_d$ are universal, that is, they are
independent on the details of the random force $f$, except via the
force length scale $L$. On top of it, since any random Gaussian
force leads to \rfs{RHS}, one could ask if at least for the
Gaussian force, the entire functions $P(\sigma)$ and $P_d(v)$ are
universal (that is, independent of the force correlation function
$F(x)$).

Some time later, an important paper by W. E, K. Khanin, A. Mazel
and Y. Sinai \cite{Sinai} came out. It was argued in that paper
that the asymptotics of $P(\sigma)$ at large negative $\sigma$ is
\be \label{LHS} P(\sigma) \propto {1 \over |\sigma|^{7 \over 2}},
\ \sigma \ll - \frac{\Urms}{L}. \ee The arguments leading to
Eq.~\rf{LHS} seemed to be well defined and valid independent of
the details of the random force setup. Those arguments were later
elaborated in a later paper by U. Frisch, J. Bec and B. Villone
\cite{Frisch}. In all of these papers, the asymptotics \rfs{RHS}
was not disputed.

At the same time, these authors argued that the only universal
feature of the \PDF $P(\sigma)$ is its ``left tail'' \rfs{LHS}.
The rest of the function was proposed to be nonuniversal,
depending on the details of the random force.

These claims were later checked numerically by R. Kraichnan and T.
Gotoh \cite{KG}, but they were not able to unambiguously confirm
or reject the power law \rf{LHS}. Some time later J. Bec
\cite{Bec} did a very thorough numerical analysis of the ``left
tail" and concluded that the power law Eq.~\rf{LHS} is indeed
correct.

Here I would like to report a numerical confirmation of the tail
\rfs{LHS}. However, I would also like to suggest that the left
tail is not the only universal feature of the function
$P(\sigma)$.

\section{Discrete Burgers Equation}

To study the Burgers equation numerically, its space and time have
to be discretized. One way to do that would be to write
derivatives as differences. That is a very crude scheme, however.
Burgers equation allows a much more gentle way of discretizing its
time, which is based on the Hopf-Cole transformation. In this
section I am going to present the scheme as it is, and show how it
produces the Burgers equation in the continuum limit.

Consider the {\sl energy function} $E_i(x)$ where $x$ changes from
$0$ to $2 \pi$ and $i$ is the integer index (which will later
become time). Now write down the following definition \be
\label{KPZ} E_i(x) = \min_{0\le y \le 2 \pi} \left[ 1-\cos
\left(x-y \right) +E_{i-1} (y) \right] + h_i(x). \ee Here $h_i$ is
a random function, which I take as \be \label{force} h_i(x) = A_i
\cos \left( x-\phi_i \right), \ee where $A_i$ is a random variable
(for example, random Gaussian with or taking values uniformly
distributed from $0$ to some maximum $A_{\rm max}$), and $\phi_i$
is a random phase uniformly distributed from $0$ to $2 \pi$. The
sign $\min$ in \rfs{KPZ} should be understood as a minimization
over $y$ which varies from $0$ to `$2 \pi$ of the expression in
the square brackets. \rfs{KPZ} plays the role of a definition of
$E_i$ in terms of $E_{i-1}$. As a starting point, one can take
$E_0(x)=0$. Finally, the functions $E_i(x)$ will automatically be
periodic in $x$.

The equation \rfs{KPZ} is the discrete Burgers equation. To see
that, let us take the continuum limit. In the notations of
\rf{KPZ}, the continuum limit is achieved when $h_i \ll 1$, $E_i
\ll 1$. The minimization condition reads \be \sin(x-y)- \partial_y
E_{i-1} (y) = 0. \ee Since $E_i$ is small, one can expand \be
\label{lagr} x = y + \partial_y E_{i-1}. \ee Substituting it back
to \rf{KPZ} and expanding $E_{i-1}(x+y-x)$ in powers of $y-x$, one
finds \be E_i(x) - E_{i-1} (x) + \oh \left( \partial_x E_{i-1}(x)
\right)^2 = h_i(x). \ee The final step is replacing the discrete
index $i$ with the continuous time $t$, to find \be
\partial_t E + \oh \left( \partial_x E \right)^2 = h,
\ee
which is the so-called KPZ equation. Introducing the velocity $u=\partial_x E$, one gets the
Burgers equation
\be
\label{bur}
\partial_t u + u~\partial_x u = \partial_x h,
\ee with $\partial_x h$ identified as the force $f$. In this
equation, the viscosity term $\nu u_{xx}$ is absent. But it is
easy to see that find a global minimum in \rf{KPZ} is equivalent
to the infinitesimal viscosity term in \rf{bur}, following, for
example, Feigelman \cite{Feigelman}. The derivation below is
somewhat technical and is given for completeness only, so it is
possible to skip the equations Eqs.~\rf{xx1}-\rf{deri} without
losing any essential information.

Introduce an alternative way of writing down \rf{KPZ}
\begin{widetext}
\be
\label{xx1}
\exp \left( - E_i(x_i) \right) = \lim_{\nu \rightarrow 0}
\int \prod_{j<i} dx_j~\exp \left\{ - \frac{2}{ \nu} \sum_{j\le i} \left[ 1-\cos \left(x_j-x_{j-1}\right)
 + h_{j}(x) \right] \right\}.
\ee
\end{widetext}
It is clear that the integral, in the limit of vanishing $\nu$
effectively minimizes the expression in the exponential, doing the
work which is accomplished by the $\min$ sign in \rf{KPZ}. Passing
to the continuum limit, one finds the functional integral
\begin{widetext}
\begin{equation}
\exp \left( -E(y,T) \right) = \lim_{\nu \rightarrow 0} \int {\cal
D} x(t) ~\exp \left\{ - \frac{2}{\nu} \int_0^T dt \left[ \oh \dot
x^2 + h(x,t) \right] \right\},  \ x(T)=y.
\end{equation}
\end{widetext}
But the right hand side is none other than the propagator of
the Schr\"odinger equation, which enables us to conclude that
$\exp \left(-E\right)$ satisfies the Schr\"odinger equation and as
a consequence, \be \label{deri}
\partial_t E + \oh \left(\partial_x E \right)^2 - \nu \partial^2_x E = h,
\ \partial_t u + u~\partial_x u - \nu \partial^2_x u = \partial_x
h, \ee which concludes the derivation. (In the last line of
\rf{deri} I replaced the arguments of the function $E(y,T)$ by $x$
and $t$.)

All this suggests that the following steps must be taken to
simulate the Burgers equation numerically. First, generate the
random ``force'' $h_i(x)$ using \rf{force}. $A_i$ can be taken,
for example, uniformly distributed on the interval from $0$ to
$A_{\rm max}$. Then solve the relationship \rf{KPZ} recursively to
find $E_i(x)$. The actual value of $E_i$ is not important for the
numerics of the Burgers equation. What is important is the value
of $y$ which minimizes the expression in the square brackets of
\rf{KPZ} for each value of $x$. Given the relationship between $x$
and $y$, one finds for the Burgers velocity $u_i = \partial_x
E_i(x)$ \be \label{burd} u_i(x) = u_{i-1}(y) + \partial_x
h_{i}(x). \ee This is the discrete analog of the obvious equation
$\dd{t} u(x(t),t) = f(x,t)$ where $\dd{t} x(t)= u(x(t),t)$, the
so-called Lagrangian trajectory (see below). Finally, for the
velocity gradients $\sigma_i(x) = \partial^2_x E_i (x)$ one finds
\be \label{sid} \sigma_i (x) = \cos\left(x-y\right)
\frac{\sigma_{i-1}(y)} {\cos \left(x-y \right) + \sigma_{i-1}(y)}
+ \partial^2_x h_{i}(x), \ee which is the discrete version of the
\rfs{Lan} considered below. The relations \rfs{KPZ}, \rfs{burd},
and \rfs{sid} are those one has to iterate to solve the Burgers
equation numerically. To stay close to the continuum limit in
time, one has to take $A_{\rm max} \ll 1$. And finally, to do the
minimization in \rf{KPZ} in practice one has to discretize space
as well. With 2000 spacial discretization point, reliable data can
be accumulated over about 10000 steps in time, which takes about
10 minutes computer time on a Sun workstation. The result of one
such calculation, with $A_{\rm max} = 0.01$ is shown on Fig 1.

\begin{figure}[htbp]
\centerline{\epsfxsize=3in \epsfbox{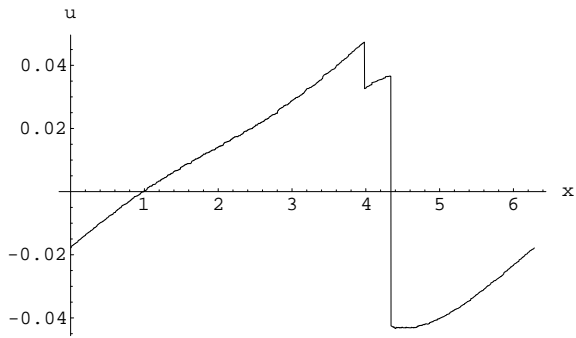}}
\caption{A typical plot of $u(x)$ at a fixed time $t$ taken from an actual computer simulation}
\end{figure}

The random force chosen according to \rf{force} fixes the function
$F$ to be $F(x) \propto \cos (x)$ and the characteristic length
scale $L$ is then of the order of the size of the system. If one
wants to use a shorter range force, one could take \be
\label{force1} h_i(x) = A_i \sum_{n=-\infty}^{\infty}\exp \left\{
- \frac{\left(x-2 \pi n+\phi_i \right)^2}{L^2} \right\}, \ee or
any other similar function, as long as it is periodic in space
$x$.

\section {Statistics of the Velocity Gradients}

The properties of the solution to Burgers equation has been discussed in many publication (see for example
\cite{Frisch}).
Here I list those which are relevant for the present discussion.

In the Burgers equation, the viscosity $\nu$ is supposed to be infinitesimally small.
Yet it cannot be set to zero. The reason for that is, the solution to this equation develop
shock waves, or sharp drops in the value of the velocity field. The width of this drops is
of the order of $1/\nu$ and they become discontinuities in $u$ when $\nu$ is taken to zero.
In the middle of the drop $\nu \partial^2_x u$ has a finite limit as $\nu \rightarrow 0$.

To study the gradients of the velocity it is advantageous to
introduce the notion of the lagrangian coordinate $x(t)$. It is
the point which moves together with the fluid \be \dbyd{x(t)}{t} =
u(x(t)). \ee A discrete version of the lagrangian trajectory was
given by \rfs{lagr}. Once $x(t)$ reaches a shock wave, it becomes
``trapped'' in it. For the purposes of this paper, it is natural
to assume that $x(t)$ ends as soon as it reaches the shock wave
(this is precisely what \rfs{lagr} does). Introduce the notion of
the gradient of the velocity on a trajectory defined by $x(t)$,
\be \label{grad1} \sigma (t) = \partial_x u(x(t),t). \ee As a
consequence of the Burgers equation, \be \label{Lan} \dot \sigma +
\sigma^2 =
\partial_x f(x(t),t). \ee Notice that the $\nu$ term can be
neglected since, by construction, $x(t)$ avoids shock waves and
terminates once it reaches them.

The equation \rfs{Lan} can be interpreted as a Langevin equation
with the random force $g(t)=
\partial_x f(x(t),t)$.
Before proceeding further, one needs to establish the statistics of $g(t)$. $x(t)$ is correlated
with $f$ in a complicated way, and it is not {\sl a priori clear} if those correlations can be neglected.

Fortunately, the discrete Burgers equation discussed in the
previous section provides an answer to this question. $g(t)$ is
none other than $\partial^2_x h_i(x)$. But $x$ is related to $y$
via \rfs{lagr}, and therefore, $x$ knows nothing about the random
phase $\phi_i$ included in the definition of $h_i$ in \rfs{force}.
Therefore, $g(t)$ is indeed just a white noise random force \be
\VEV{g(t) g(t')} =  G~\delta(t-t') \ee where, as before,
$G=-\partial^2_x F(x)|_{x=0}\approx(\Urms/L)^3$. Given a Langevin
equation \rfs{Lan}, a Fokker-Planck equation can be written down
for the propability $P(\sigma)$ \be \pp{\sigma} \left( {G \over 2}
\pp{\sigma} + \sigma^2 \right) P = \pbyp{P}{t}. \ee In the present
context this equation was first written down in \cite{Bouchaud},
but in that paper it was given an interpretation different from
the one here. In fact, a slightly more general form of this
equation will be useful below. It has the form \be \label{FP}
\pp{\sigma} \left( {G \over 2} \pp{\sigma} + \sigma^2 \right) P +
\beta P = \pbyp{P}{t} \ee where $\beta$ is an arbitrary parameter.
At this stage, $\beta=0$. However, later a version of \rfs{FP}
with nonzero $\beta$ will be useful. Notice that the large
$\sigma$ asymptotics of $P$ obtained from the \rfs{FP} is
\rfs{RHS}, irrespective of the value of $\beta$.

\rfs{Lan} is unstable. As soon as $\sigma$ becomes large negative,
it quickly reaches infinity. This process is none other than the
formation of a shock wave. A quick estimate of the probability
$P(\sigma)$ in this regime is easy to do. Once $\sigma$ is large
enough and negative, one can neglect $g(t)$ in \rfs{Lan}
completely and solve the equation to find \be \sigma(t) =
\frac{1}{t-t_0}, \ t < t_0 \ee The probability can be estimated to
be \be \label{naive} P(\sigma) \propto \int
dt~\delta(\sigma-\sigma(t)) \propto \frac{1}{\sigma^2}, \ \sigma
\ll -\frac{\Urms}{L}. \ee This estimate was used in some
publications to conclude that this should be the correct power law
in the \PDF of the Burgers equation \rfs{LHS}. However, to say so
would be premature.

It is instructive to reproduce the power law \rfs{naive} with the
help of the Fokker-Planck equation. To do that, observe that one
has to solve this equation with an absorbing boundary conditions
at $\sigma \rightarrow -\infty$, corresponding to the trajectories
disappearing when shock waves are formed. The standard
substitution \be P=\exp\left( - \frac{\sigma^3}{3 G} \right)
\Psi(\sigma). \ee leads to the Schr\"odnger equation of a quantum
mechanical particle moving in a potential $U(\sigma) = \sigma^4/(2
G) - (\beta+1)\sigma$ \be \label{eq1} -\frac{G}{2}
\frac{\partial^2 \Psi}{\partial \sigma^2} +
\left(\frac{\sigma^4}{2 G} - \left( \beta + 1 \right) \sigma
\right) \Psi = - \pbyp{\Psi}{t}. \ee The solution to this equation
reads \be \Psi(\sigma) = \sum_{n=0}^{\infty} C_n \exp \left( - E_n
t \right) \Psi_n(\sigma), \ee where $C_n$ are arbitrary
coefficients, $E_n$ are the eigenstates, and $\Psi_n$ are the
eigenfunctions of the Schr\"odinger equation satisfying \be
-\frac{G}{2} \frac{\partial^2 \Psi_n}{\partial \sigma^2} +
\left(\frac{\sigma^4}{2 G} - \left( \beta+1 \right) \sigma \right)
\Psi_n = E_n \Psi_n. \ee At large times $t \gg (E_1-E_0)^{-1}$,
only the ground state survives
\begin{eqnarray}
\label{eq2}&  P(\sigma,t) \propto \exp \left( - E_0 t \right)
P_0(\sigma), \ t \gg \left(E_1-E_0\right)^{-1}, \cr & P_0(\sigma)
= \exp\left( - \frac{\sigma^3}{3 G} \right) \Psi_0 (\sigma).
\end{eqnarray} A detailed analysis of Eqs.~\rf{eq1}-\rf{eq2} at arbitrary
$\beta$ can be found in the paper by S. Boldyrev \cite{Boldyrev}.
In the case of interest, $\beta=0$. Ref. \cite{Boldyrev} shows
that in this case $E_0>0$, and it confirms the asymptotics
\rfs{naive} for $P_0(\sigma)$. The asymptotics at large positive
$\sigma$ coincides with \rfs{RHS}. The fact that $E_0>0$ is not
surprising. It means that the total probability $\int d
\sigma~P(\sigma)$ dicreases with time. This is due to elimination
of trajectories which ended on shock waves.

However, this cannot be the correct probability of observing a velocity gradient $\sigma$ at a given
point in space. Indeed, introduce the density of trajectories $\rho(x,t)$. It satisfies
the continuity equation
\be
\partial_t \rho + \partial_x \left(u \rho \right) = 0.
\ee In particular, the density around the given Lagrangian
trajectory $x(t)$ is given by $\rho_L(t) = \rho(x(t),t)$ and it
satisfies \be \dd{t} \rho_L + \sigma \rho_L = 0 \ee It is clear
that the probability that a particular trajectory happens to go
through an observation point where a velocity gradient is measured
is inversely proportional to the density of trajectories at this
point. Thus a more physically relevant probability distribution
function is defined as \be P(\sigma) = \VEV{ \frac{1}{\rho_L}
~\delta (\sigma-\sigma(t))}. \ee This function satisfies a
Fokker-Planck equation \rfs{FP}, but with $\beta=1$. In this case,
it is possible to show that $E_0<0$ and \be P(\sigma) \propto
\frac{1}{|\sigma|^3}, \ \sigma \ll -\frac{\Urms}{L}. \ee This
asymptotics was proposed in the paper \cite{KG}. But it cannot be
the correct asymptotics either. Indeed, while the above analysis
correctly removes the trajectories which end at the points where a
new shock wave is formed, it fails to take into account those
which fall into ``mature'' shock waves, the ones which formed in
the past. That is why $E_0<0$ and the probability grows with time.
That comes from overcounting the trajectories.

Unfortunately, the \rfs{Lan} does not seem to contain enough
information to determine if there is a shock wave nearby. If that
is indeed the case, then the full Burgers equation \rfs{burgers}
has to be solved. That is the core of the problem of Burgers
turbulence, and the reason the solution to it seems so close and
yet so hard to get. The full Burgers equation \rfs{burgers} leads
to a full fledged 2D quantum field theory (via Martin-Siggia-Rose
formalism, for example). That theory is presumably very hard, if
not impossible, to solve. The ``reduced'' equation \rfs{Lan}, on
the other hand, leads to a quantum mechanics encoded in \rfs{eq1}.
The solution to quantum mechanics problems are clearly within
reach. The question is, therefore, if it is possible to extract
all the relevant information about the Burgers equation from the
Lagrangian equation \rfs{Lan}. At this stage of the present
analysis, it does not seem possible.

To move further, I will make the following leap of faith. The
equation \rfs{FP} was studied in Ref.~\cite{Boldyrev} for
arbitrary $\beta$. It was found that while the large positive
$\sigma$ asymptotics of $P$ is always given by \rfs{RHS}, the
other asymptotics is given by \be P(\sigma) \propto  \frac{1}
{\sigma^{\beta+2}}, \ \sigma \ll -\frac{\Urms}{L}. \ee On the
other hand, it is known from the analysis of Ref.~\cite{Sinai} and
\cite{Frisch} that the correct power of this asymptotics is $7/2$.
To match this asymptotics, I choose $\beta=3/2$. Then I propose to
solve the equation \rfs{FP} and compare what it gives with the
numerics. The results of the comparison is shown in next section.

The equation \rfs{FP} with an arbitrary $\beta$ was proposed in
\cite{Polyakov} (but for $P_d(v)$ as opposed to $P(\sigma)$) where
$\beta$ was referred to as $b$-anomaly. But whether the present
discussion has anything to do with the methods of
Ref.~\cite{Polyakov} is unclear. To derive \rfs{FP} with
$\beta=3/2$, one needs to show that the rate at which the
trajectories fall into mature shock waves is proportional to the
extra $\oh \sigma P$ term generated in the Fokker-Planck equation
\rfs{FP}. How to do that is not known to me. Next, the \rfs{FP}
cannot be exact. After all, it clearly breaks down even if the
force is not Gaussian. However, it may be correct in some mean
field sense, for moderate values of $\sigma$. Finally, I'd like to
note that for this value of $\beta$, $E_0<0$ and the probability
still grows with time. That must be because this description gives
the relative probability of observing one value of $\sigma$ over
another, not the absolute probability, and the overall
normalization must be enforced by $P \rightarrow P/\Pi$ where
$\Pi=\int d\sigma~P(\sigma)$.

\section{Numerical Simulations}
With the setup descibed in the preceeding sections, it is very
easy to calculate $P(\sigma)$ numerically. Unlike some of the
previous work on this subject, I did not try to compute
$P(\sigma)$ from the data. Instead, I computed the integrated
probability density \be N(\sigma) = \int_{-\infty}^{\sigma}
d\mu~P(\mu). \ee Obviously its left tail is expected to go as
$N(\sigma) \propto 1/|\sigma|^{\frac{5}{2}}$. To plot it
numerically, it is enough to collect the data into an array, to
sort it in the increasing order, and plot the position of an array
entry as a function of its value. Fig 2. depicts $N(\sigma)$ in
the case when the force was chosen to be as in \rfs{force},
$A_{\rm max}=0.01$, and $\Urms \approx 0.02$.
\begin{figure}[htbp]
\centerline{\epsfxsize=3in \epsfbox{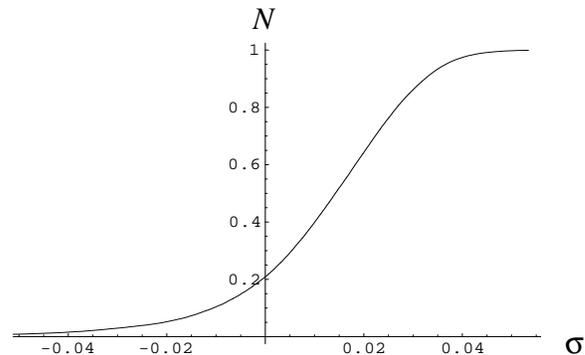}}
\caption{The integrated probability density $N(\sigma)$ as a function of $\sigma$}
\end{figure}

Fig 3. shows the left tail in
the log-log format. For comparison, a straight line with the slope $-5/2$ is plotted.
\begin{figure}[htbp]
\centerline{\epsfxsize=3in \epsfbox{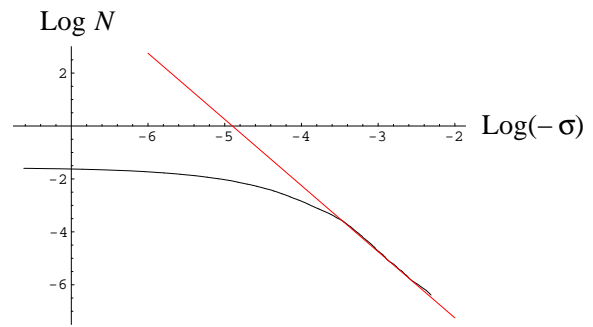}}
\caption{The plot of $\log N$ as a function of $\log(-\sigma)$ for negative $\sigma$. The straight
line has a slope of $-2.5$}
\end{figure}
The agreement is striking. One could object that the power law in
this graph does not extend over a sufficiently long range of
$\sigma$. To extend it further, one needs to do numerics at weaker
force for longer periods of time (to keep close to the continuum
limit even at large velocity fluctuations). It was not the task of
the present work to do large scale numerical simulations, and I
believe the agreement I have here is good enough (especially in
view of the work reported in Ref.~\cite{Bec}).

Fig 4. shows the same probability function, but plotted together
with the (numerical) solution to the equation \rfs{FP} with
$\beta$ chosen to be $\beta=3/2$.
\begin{figure}[htbp]
\centerline{\epsfxsize=3in \epsfbox{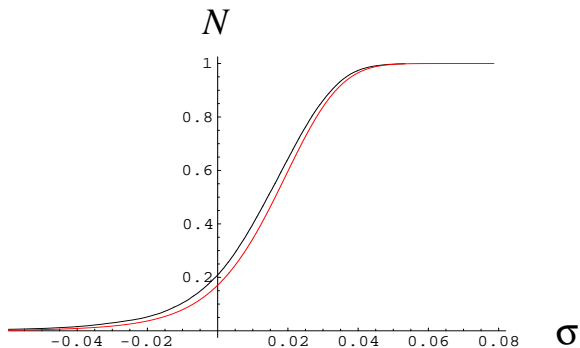}}
\caption{$N(\sigma)$ taken from the numerics (upper curve) and
extracted from the solution to the \rfs{FP} with $\beta=3/2$.}
\end{figure}
To arrive at this picture, one parameter - the units of measure of
$\sigma$ - had to be adjusted to make the graphs overlap as much
as possible. The agreement is not perfect. Yet the curves are
close enough to suggest that perhaps it is not coincidential
either. One interesting number is the probability $p$ that the
velocity gradient is negative. It is especially interesting,
because it is independent of the adjustments of the units of
$\sigma$. Numerically $p\approx 0.21\pm 0.01$. The solution to the
\rfs{FP} gives about $0.18$. According to some claims in the
literature, this number is not universal. However, the value of
$p$ is reproducible whether I do the numerics with the long ranged
force \rfs{force} or shorter range force \rfs{force1} with various
values of $L$. So numerically $p$ appears to be universal, or
perhaps approximately universal.

To conclude, the numerical data does not confirm that the \rfs{FP}
with the parameter $\beta=3/2$ is the {\sl exact} solution to the
problem. But it may be an {\sl approximate}, perhaps in some mean
field sense, solution to the problem, which reproduces the actual
probability density rather close. I believe this question deserves
further investigations.

\section{Pinned Charge Density Waves}

In the papers of the author with J.T. Chalker \cite{GC,GC1} a
remarkable correspondence was established between the randomly
driven Burgers equation and the physics of pinned charge density
waves. It would go beyond this paper to discuss charge density
waves, but mathematically the correspondence can be formulated in
the following way. Consider the functional
\begin{equation}
\label{PCDW} {\cal E}[x(t)] = \int_0^T dt \left[ \oh \dot x^2 +
h(x,t) \right],
\end{equation}
where $h(x,t)$ is the same random function as the one in
\rfs{xx1}. Let us find a function $x_0(t)$ which, when substituted
for $x(t)$ in \rfs{PCDW}, gives an absolute minimum for the
functional ${\cal E}$. Obviously it satisfies the minimization
equation
\begin{equation}
-{d^2 x_0\over dt^2} + \left.\pbyp{h(x,t)}{x}\right|_{x=x_0(t)} =
0.
\end{equation}

Now consider the energy cost of $x(t)$ deviating from $x_0(t)$.
Writing $x(t) \approx x_0(t) + \psi(t)$ we find the equation for
the normal modes of oscillations about the absolute minimum, given
by
\begin{equation}
\left[ -{d^2  \over dt^2}+ \left. {\partial^2 h(x,t) \over \d x^2}
\right|_{x=x_0(t)} \right]\psi = \omega^2 \psi.
\end{equation}
With the problem thus set up, one needs to calculate the average
number of modes with frequencies less than a given frequency
$\omega$. Such a function is denoted as $N(\omega)$.

This problem, which was first formulated in the context of charge
density waves in Ref.~\cite{FL}, received a substantial amount of
attention in the literature (see
Refs.~\cite{Feigelman,Gia,AR,Fogler,GC,GC1}. After initial
disagreements, a consensus was built which gave for the function
$N(\omega)$ the value
\begin{equation}
\label{IDS} N(\omega) \propto \omega^5,
\end{equation}
for sufficiently small $\omega$. This was first proposed in
Ref.~\cite{AR} and then the derivation was improved in subsequent
publications until Ref.~\cite{GC} derived \rfs{IDS} in a
systematic way.

One of the results of Refs.~\cite{GC,GC1} relates the calculation
of $N(\omega)$ to the following specific question in Burgers
turbulence. Consider a Burgers fluid \rfs{burgers} driven by a
random force $f(x,t)=\partial_x h(x,t)$. Consider a Lagrangian
trajectory $x_0(t)$ which moves with a fluid and never gets
absorbed by a shock wave for times $t<T$. Consider a velocity
gradient $\sigma(t)$ defined in \rfs{grad1}. Let us find time
intervals $t_1 < t < t_2$ such that $\sigma(t)<0$ within those
intervals. Now let us calculate the probability that
\begin{equation}
\label{cond1} \int_{t_1}^{t_2} dt~\sigma(t) < \log(\omega),
\end{equation}
for some small value of $\omega$. Then the function $N(\omega)$ of
the charge density wave problem coincides with this probability.

Notice that knowing the tails of the probability distribution
$P(\sigma)$ would not help calculating $N(\omega)$. This is
because the typical functions $\sigma(t)$ which satisfy
\rfs{cond1} are not those for which $\sigma$ is very large
negative, but rather moderate negative $\sigma(t)$ which however
persist over long time intervals. Therefore, the results of W. E
{\sl at al} \cite{Sinai} are useless if we were to calculate
$N(\omega)$ with the help of \rfs{cond1}.

Let us however use the methods of this paper, in particular
\rfs{eq1}, to calculate $N(\omega)$. To do so, we need to write
down a Feynman path integral formalism which corresponds to
\rfs{eq1} and then use the Lagrange multiplier method described in
Ref.~\cite{GC1}, section VI C, subsection 1, corrected however for
the presence of a parameter $\beta$ in \rfs{eq1}. Then we find
\begin{equation}
N(\omega) \propto \omega^{2 \beta+2}.
\end{equation}
Now we know the exact answer to the problem, \rfs{IDS}. If the
methods described in this paper were to reproduce the exact
answer, we have to choose $\beta=3/2$. However, {\sl this is the
same value of $\beta$ required to reproduce the asymptotics of W.
E {\sl et al}!}

Two possible explanations of this are possible. First is, perhaps
the \rfs{eq1} with $\beta=3/2$ indeed reproduces the right
asymptotics of W. E {\sl et al} and for some reason also
reproduces the function $N(\omega)$ correctly, but this equation
is still generally not correct and the probability distribution
derived with its help and shown on Fig. 4 is not correct either.
Second is, the \rfs{eq1} is indeed correct and gives the correct
solution to the Burgers problem.

The first explanation seems far fetched, as a remarkable
coincidence must be at play to give rise to it. The second
explanation sounds much more likely. And yet, no derivation of
\rfs{eq1} is known at this point.

\section {Conclusion}
In this paper, I demonstrate that it is easy to obtain reliable
numerical data on the behavior of the 1D Burgers equation. I show
that the data confirms the $7/2$ tail of the probability
distribution suggested in the literature. The data also suggests
that the entire probability distribution function could be
universal, determined by the solution to the equation \rfs{FP}
with the ``anomaly'' parameter $\beta=3/2$. I also discuss that it
is possible to reproduce the known solution to the pinned charge
density wave problem by applying the techniques discussed here and
choosing $\beta=3/2$, which gives extra weight to these methods.

\section{Acknowledgements}
I am grateful to J.T. Chalker for many discussions. This paper is
based on the results obtained in part in the course of work on
paper \cite{GC}. The work presented here was supported in part by
EPSRC through its Advanced Research Fellowship programme.

\end{document}